# Robustness of the second law of thermodynamics under generalizations of the maximum entropy method

S. ABE[1,2] and S. THURNER[3,4]

[1]*Department of Physical Engineering, Mie University, Mie 514-8507, Japan*

[2]*Institut Supérieur des Matériaux et Mécaniques Avancés, 44 F. A. Bartholdi, 72000 Le Mans, France*

[3]*Complex Systems Research Group, HNO, Medical University of Vienna, Währinger Gürtel 18-20, A-1090, Austria*

[4]*Santa Fe Institute, 1399 Hyde Park Road, Santa Fe, NM 8750, USA*

**Abstract** — It is shown that the laws of thermodynamics are extremely robust under generalizations of the form of entropy. Using the Bregman-type relative entropy, the Clausius inequality is proved to be always valid. This implies that thermodynamics is highly universal and does not rule out consistent generalization of the maximum entropy method.

There is a consensus of a certain kind that among the physical laws those of thermodynamics are extremely universal and will persist. This optimism might have its origin in the fact that the thermodynamic principles have remained unchanged when the shift was made from classical statistical mechanics to quantum. This is indeed astonishing if one takes into account that the formulation of thermodynamics was completed in the 19th century, before the establishment of atomism in physics. Here, by the term "statistical mechanics", we refer to ordinary equilibrium statistical mechanics that describes normal systems, in which ergodicity is supposed to be well realized (though it still seems vague in quantum theory).

Today, it is well known that there exist a number of systems in nature, which are essentially nonequilibrium and nonergodic, but exhibit definite statistical properties. In particular, nonequilibrium stationary states of complex systems are of primary interest in various branches of physical science. These systems are often characterized by nonexponential distributions including the power-law type. An important point here is that in many such systems the stationary states survive for very long periods of time, much longer than typical microscopic dynamical time scales. This makes it legitimate to suppose that such a state may maximize a certain statistical measure, which can be different from the Boltzmann-Gibbs entropy, in general. Here is a point, at which generalization of the ordinary maximum entropy method is expected to play a role.

Now, the above-mentioned reasoning brings thermodynamics a new issue to be examined. That is, how robust the laws of thermodynamics are under such a generalization. The purpose of the present work is to answer this question. We execute it by employing the highly generalized entropic measure proposed recently and examining mainly the second law for the generalized maximum entropy method based on it. We shall see how the Clausius inequality is robust under such a generalization.

There are several works devoted to highly generalized entropic measures in the literature [1-6]. Here, let us employ the one recently proposed in Ref. [5], which may be the most general form that is consistent with the maximum entropy condition. In the unit in which the Boltzmann constant is unity, it reads

$$S_G[B] = -\sum_i B(\varepsilon_i) \Lambda(B(\varepsilon_i)) + \eta[B] \tag{1}$$

with

$$\eta[B] = \sum_i \int_0^{B(\varepsilon_i)} dx\, x\, \frac{d\Lambda(x)}{dx} + c, \tag{2}$$

which can be rewritten as

$$S_G[B] = -\sum_i \int_0^{B(\varepsilon_i)} dx\, \Lambda(x) + c. \tag{3}$$

In these equations, $B(\varepsilon_i)$ is a normalized and monotonically decreasing distribution function of $\varepsilon_i$, which is the value of energy of the system in its $i$th state. $\Lambda(x)$ is a "generalized logarithm" satisfying $\Lambda(x) < 0$ and $d\Lambda(x)/dx > 0$ for $0 < x < 1$, and $x\Lambda(x) \to 0$ ($x \to 0+$), making $S_G[B]$ nonnegative and concave (its further properties will be specified later). $c$ is a constant, which ensures that $S_G[B_0] = 0$ for a completely ordered state, $B_0[\varepsilon_i] = \delta_{\varepsilon_i, \varepsilon_0}$ with the "ground-state energy" $\varepsilon_0$. This condition is necessary for the third law to hold. Note however that the third law is fundamentally concerned with quantum theory in the low temperature regime, whereas we are considering a classical system here.

A generalized maximum entropy method, given the existence of a stationary distribution function, $\tilde{B}(\varepsilon_i)$, is formulated as follows:

$$\left. \frac{\delta G}{\delta B(\varepsilon_i)} \right|_{B=\tilde{B}} = 0 \tag{4}$$

with

$$G \equiv S_G[B] - \alpha \left\{ \sum_i B(\varepsilon_i) - 1 \right\} - \beta \left\{ \sum_i \varepsilon_i B(\varepsilon_i) - U \right\}, \tag{5}$$

where $\alpha$ and $\beta$ are Lagrange multipliers, and $U$ denotes the average energy. The stationary solution to this problem is given by

$$\tilde{B}(\varepsilon_i) = E(-\alpha - \beta(\varepsilon_i - \tilde{U})), \tag{6}$$

where $E(x)$ is a "generalized exponential", which is the inverse function of $\Lambda(x)$: $E(\Lambda(x)) = \Lambda(E(x)) = x$. $\tilde{U}$ denotes the average energy evaluated in the stationary state in Eq. (6) in a self-referential manner. In other words, $\Lambda(x)$ in Eq. (1) is chosen so that it is the inverse function of the stationary distribution function. $\alpha$ in Eq. (6) can be eliminated as follows [7]. In addition to Eq. (4), we consistently impose the following condition:

$$\sum_i B(\varepsilon_i) \frac{\delta}{\delta B(\varepsilon_i)} G \bigg|_{B=\tilde{B}} = 0. \tag{7}$$

Combining this with the normalization condition, $\sum_i \tilde{B}(\varepsilon_i) = 1$, we obtain

$$\alpha = -\beta \tilde{F} - \eta[\tilde{B}], \tag{8}$$

where

$$\tilde{F} \equiv \tilde{U} - \beta^{-1} S_G[\tilde{B}]. \tag{9}$$

On the other hand, $\beta$ is found to satisfy

$$\frac{\partial S_G[\tilde{B}]}{\partial \tilde{U}} = \beta, \tag{10}$$

which is actually nothing else than a generic feature in the variational problem in Eqs. (4) and (5) [8]. Thus, the thermodynamic Legendre transformation structure always exists, and accordingly $\tilde{F}$ in Eq. (9) is regarded as the free energy in the stationary state.

It should however be noticed that the zeroth law of thermodynamics does not hold in the present case, since it is not possible to divide the total system into two independent subsystems and the entropic measure is not additive, in general. This makes it hard to interpret $\beta^{-1}$ as a "temperature" (recall that the temperature specified by the zeroth law is an equilibrium temperature). Taking into account the physical properties of a nonequilibrium stationary state, it is natural to regard $\beta^{-1}$ as a *local* temperature that may vary both spatially and temporally.

The first law of thermodynamics reflects the conservation of energy and is therefore valid also in any nonequilibrium situation. Let us look at the definition of the internal energy in a generic state, $B(\varepsilon_i)$

$$U = \sum_i \varepsilon_i B(\varepsilon_i). \qquad (11)$$

Change in a thermodynamics process is

$$\delta U = \sum_i \varepsilon_i \delta B(\varepsilon_i) + \sum_i B(\varepsilon_i) \delta \varepsilon_i, \qquad (12)$$

where $\delta B(\varepsilon_i)$ implies the change of the functional form of $B(\varepsilon_i)$ whereas $\delta \varepsilon_i$ is the change of the energy due to a certain external influence. Identifying the quantity of heat and the work respectively as

$$\delta' Q = \sum_i \varepsilon_i \delta B(\varepsilon_i), \qquad (13)$$

$$\delta' W = -\sum_{i} B(\varepsilon_i) \, \delta \varepsilon_i, \tag{14}$$

we immediately obtain the first law of thermodynamics

$$\delta' Q = \delta U + \delta' W. \tag{15}$$

To ascertain that the above identifications are legitimate, let us consider a "quasi-reversible process" [9] connecting two stationary states that are infinitesimally distinct each other. Then, the change of $S_G$ is evaluated for the stationary state in Eq. (6) as follows:

$$\delta S_G = -\sum_{i} \Lambda(\tilde{B}(\varepsilon_i)) \delta B(\varepsilon_i)$$

$$= \beta \sum_{i} \varepsilon_i \, \delta B(\varepsilon_i), \tag{16}$$

provided that the normalization condition on $B(\varepsilon_i)$ has been used. Therefore, we find

$$\delta S_G = \beta \delta' Q. \tag{17}$$

This relation for a quasi-reversible process justifies the identifications mentioned above.

Now let us proceed to study the second law of thermodynamics. For this purpose, we consider comparing a general distribution function, $B(\varepsilon_i)$, with the stationary state distribution function, $\tilde{B}(\varepsilon_i)$. In information theory, this can traditionally be done by making use of the Kullback-Leibler relative entropy (or, divergence) [10,11], which is intrinsically associated with the ordinary Boltzmann-Gibbs-Shannon entropic measure. However, it is a nontrivial problem in our case, since we are concerned with a highly generalized form of entropy in Eq. (1).

For a solution to this problem, a clue may be found in the work in Ref. [7]. There, it is indicated that the relative entropy associated with the ordinary definition for expectation values is the one of the Bregman type [12] (see also Ref. [3]). In the present context, the relative entropy of this type reads

$$K_G[B\|\tilde{B}] = \sum_i \int_{\tilde{B}(\varepsilon_i)}^{B(\varepsilon_i)} dx\, [\Lambda(x) - \Lambda(\tilde{B}(\varepsilon_i))], \tag{18}$$

which can be recast in the form

$$K_G[B\|\tilde{B}] = S_G[\tilde{B}] - S_G[B] + \beta U - \beta \tilde{U}$$
$$\equiv \beta \Delta F \tag{19}$$

with

$$\Delta F = F - \tilde{F} = (U - S[B]/\beta) - (\tilde{U} - S[\tilde{B}]/\beta). \tag{20}$$

Therefore, one sees that the physical meaning of $K_G[B\|\tilde{B}]$ is essentially the "free energy" difference.

Let us recall some fundamental properties of $K_G[B_1\|B_2]$. The following two are of importance for our discussion. One is positive semi-definiteness: that is, $K_G[B_1\|B_2] \geq 0$ and $K_G[B_1\|B_2] = 0$ iff $B_1(\varepsilon_i) = B_2(\varepsilon_i)$. The other is its convexity with respect to the first argument:

$$K_G[\lambda B_1 + (1-\lambda) B_2 \| B_3] \leq \lambda K_G[B_1 \| B_3] + (1-\lambda) K_G[B_2 \| B_3], \tag{21}$$

where $\forall \lambda \in (0, 1)$.

Now, in connection with Eq. (18), take some arbitrary third state, $B^*$, satisfying

$$K_G[B^* \| \tilde{B}] \leq K_G[B \| \tilde{B}]. \tag{22}$$

Then, define the variation of the state $B(\varepsilon_i)$ naturally as follows:

$$\delta B(\varepsilon_i) = \{\lambda B^*(\varepsilon_i) + (1-\lambda)B(\varepsilon_i)\} - B(\varepsilon_i). \tag{23}$$

Under this variation, the change of $K_G[B \| \tilde{B}]$ with fixed $\tilde{B}(\varepsilon_i)$ yields

$$\delta_B K_G[B \| \tilde{B}] = \sum_i [\Lambda(B(\varepsilon_i)) - \Lambda(\tilde{B}(\varepsilon_i))]\delta B(\varepsilon_i)$$

$$= \sum_i \Lambda(B(\varepsilon_i))\delta B(\varepsilon_i) + \beta \sum_i \varepsilon_i \delta B(\varepsilon_i). \tag{24}$$

That is,

$$\delta_B K_G[B \| B'] = -\delta S_G + \beta \delta' Q. \tag{25}$$

From Eqs. (22) and (23), we see that the left-hand side of this equation is not positive:

$$\delta_B K_G[B \| \tilde{B}] \equiv K_G[\lambda B^* + (1-\lambda)B \| \tilde{B}] - K_G[B \| \tilde{B}]$$

$$= \lambda \{K_G[B^* \| \tilde{B}] - K_G[B \| \tilde{B}]\} \leq 0. \tag{26}$$

Consequently, we obtain the Clausius inequality

$$\beta \delta' Q \leq \delta S_G. \tag{27}$$

In conclusion, we have shown that the laws of thermodynamics remain fundamentally unchanged under generalizing the form of entropy in the maximum entropy method. In particular, the second law, expressed in the form of the Clausius inequality does not change for a very wide class of generalized entropies. Thus, thermodynamics is highly universal and does not rule out any effort for generalizing the form of entropy.

* * *

The work of S. A. was supported in part by Grant-in-Aid for Scientific Research (B) of the Ministry of Education. S. T. is grateful to Austrian Science Foundation Projects P17621 and P19132.